\documentclass[reprint,amsmath,amssymb,pra]{revtex4-1}

\usepackage{graphicx}
\usepackage{color}
\usepackage[dvipsnames]{xcolor}
\usepackage{colortbl}
\usepackage{import}
\usepackage[cp1252]{inputenc}
\usepackage{dcolumn}
\usepackage{bm}
\usepackage{placeins}
\usepackage{float}
\usepackage{subfigure}
\usepackage{subfig}
\usepackage{braket}
\usepackage{mathdots}
\usepackage{verbatim}
\usepackage{subfigure}
\usepackage{ulem}
\usepackage{tikz}
\usepackage{amsmath}
\usepackage{filecontents}

\newcommand{\e}{\mathrm{e}}

\begin{document}

\title{
Quantum gravimetry in the same internal state using composite light Raman pulses 
}

\author{G.~A. Olivares-Renter\'ia}
\affiliation{Instituto de F\'isica, Universidad Aut\'onoma de San Luis Potos\'i, San Luis Potos\'i, 78290, Mexico}
\affiliation{Byteflow Dynamics LLC, New York, 11103, USA}
 
\author{D.~A. Lancheros-Naranjo}
\affiliation{Instituto de F\'isica, Universidad Aut\'onoma de San Luis Potos\'i, San Luis Potos\'i, 78290, Mexico}%

\author{E. Gomez}
\affiliation{Instituto de F\'isica, Universidad Aut\'onoma de San Luis Potos\'i, San Luis Potos\'i, 78290, Mexico}%

\author{J.~A. Franco-Villafa\~ne}
\email{jofravil@ifisica.uaslp.mx}
\affiliation{CONACYT - Instituto de F\'isica, Universidad Aut\'onoma de San Luis Potos\'i, San Luis Potos\'i, 78290, Mexico}%

\date{\today}

\begin{abstract}
We introduce an atomic gravimetric sequence using Raman-type composite light pulses that excites a superposition of two momentum states with the same internal level. The scheme allows the suppression of common noise, making it less sensitive to external fluctuations of electromagnetic fields. The Raman beams are generated with a fiber modulator and are capable of momentum transfer in opposite directions. We obtain analytical expressions for the interference fringes in terms of three perturbative parameters that characterize the imperfections due to undesired frequencies introduced by the modulation process. We find special values of the Rabi frequency that improve the fringes visibility.
\end{abstract}

\maketitle

\section{Introduction}

\label{intropaper}
State of the art gravimeters achieve a sensitivity of about $10^{-8}$m/s$^2$. Better sensitivity or faster measurements would benefit some of their applications, such as oil exploration or early detection of earthquakes~\cite{Montagner2016}, among others. Atomic gravimeters have overcome some of the limitations of classical gravimeters and they already show better sensitivity \cite{Menoret2018}.

One widespread implementation makes use of Raman transitions, which require two counter-propagating lasers separated in frequency by the atomic hyperfine splitting \cite{MeasurementK}. Traditionally the two frequencies are obtained from independent lasers, that are referenced with respect to each other by a phase lock loop ~\cite{Yim,Gouet,Cheinet,Santarelli,Cacciapuoti,Gouet2}. Alternatively, it is possible to use a light modulator to introduce sidebands that work as the additional frequencies ~\cite{Dickerson,JWang, Doring, QWang, JLee,KLee,Charriere,Arias17}. Since they come from a single laser, this method gives very low phase noise ~\cite{Cheinet2,Carraz1,Theron,Bidel}.

The Raman pair (carrier and +1 sideband for example) prepares a coherent superposition of $\left| a,p \right\rangle$ and $\left| b,p+ 2\hbar k \right\rangle$ with $a$ and $b$ two hyperfine levels and $p$ the momentum. Having a third frequency at the modulator output (-1 sideband in this case) gives additional contributions that have to be taken into account \cite{Carraz2012}. For highly detuned co-propagating transitions from phase modulators, there is destructive interference between the two Raman pairs generated \cite{JLee, Debs}. This problem can be avoided by eliminating the additional frequency using serrodyne modulation to generate only a single sideband when there is enough bandwidth available \cite{Johnson}, by using narrow optical filters to eliminate a particular frequency \cite{KLee}, or by using birefringent dispersive materials \cite{Arias17}.

Here we use this third frequency to our advantage to enable momentum transfer in two opposing directions. The appropriate frequency tuning selects the direction of momentum transfer. The excitation scheme produces a superposition between different momentum states with the same internal level through a tailored sequence of Raman pulses. A similar technique has been used to measure the Earth's rotation rate with symmetric momentum-space splitting~\cite{Berg2015}. Since the atoms remain in the same internal state during the dark periods, the interferometer is more robust against environmental fluctuations.
The scheme is less sensitive to many systematic effects, such as AC-Stark or Zeeman shifts, just as with Bragg diffraction atomic interferometers~\cite{Mueller2009,Leveque2009,Altin2013,Ahlers2016}, but keeping the simple spectroscopic readout of the Raman interferometers.

In this work, we derive analytical expressions for the proposed excitation sequence, and we identify three perturbative parameters that characterize the effect of the imperfections. The article is organized as follows. Section \ref{Intvel} presents the scheme for atomic gravimetry in a momentum superposition. Section \ref{Dynamics} introduces the Hamiltonian of the Raman excitation in the reference frame of the free-falling atom. Section \ref{DiffEqs} gives a perturbative approach to solve the atomic dynamics. Sections \ref{pulse1}, \ref{pulse2} and \ref{fringesg} show the results for the first and second pulse and for the complete interferometric sequence respectively.

\section{Interferometry in a momentum superposition}\label{Intvel}

\begin{figure}[H]
\centering
\includegraphics[width=80mm]{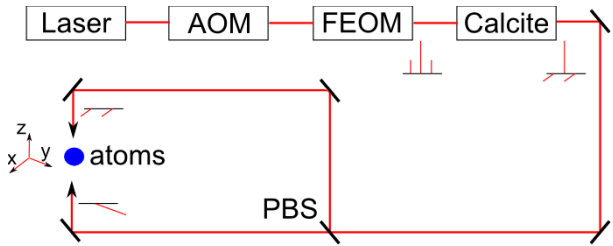}
\caption{Setup for two Raman pairs with opposite momentum transfer. AOM: acousto-optical modulator, FEOM: fiber electro-optical modulator, PBS: polarization beam splitter. The polarizations of the carrier and sidebands are also shown.}
\label{setupintvel}
\end{figure}

The proposed setup is shown in Fig.~\ref{setupintvel}. We consider the case of hyperfine transitions in alkali atoms. The light from a laser goes through an acousto-optical modulator to generate the pulses, and through a fiber phase modulator to produce the $\pm 1$ sidebands. Sending the phase modulator output light through a calcite crystal in the correct configuration rotates the polarization of the sidebands by 90 degrees leaving the carrier unchanged \cite{Arias17}. We split them by polarization to send them in opposite directions. The carrier (with wavevector $\vec{k}_c$) and the $+1$ sideband ($\vec{k}_{+1}$) provide a Raman pair that transfers momentum in one direction whereas the $-1$ sideband ($\vec{k}_{-1}$) and the carrier produce another pair with opposite momentum transfer. There are no co-propagating Raman transitions in this setup.

For an atom of mass $m$ and velocity $v_z$ in the $z$-direction, the momentum transfer of the counter-propagating beams in the $z$ axis introduces a detuning given by~\cite{Moler1992}
\begin{equation}
\label{delta}
\delta = \pm k_{e} v_z + \hbar k_{e}^2/2m,
\end{equation}
where the sign depends on what Raman pair we are considering and  $k_{e}=|\vec{k}_{\pm 1}-\vec{k}_c|$ $\approx2 k_{c}$ is the effective wave vector between the carrier and the sidebands. Since both Raman pairs have an opposite Doppler shift, we can tune each transition on resonance separately as long as the atomic velocity is not close to zero.

\begin{figure}[tbp]
\centering
\includegraphics[width=80mm]{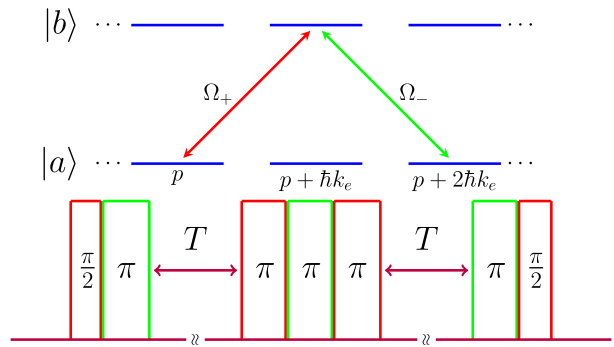}
\caption{Sequence of pulses required to excite a momentum superposition.}
\label{fig:sequencepulses}
\end{figure}

In a traditional gravimeter, one applies a sequence of three pulses ($\pi/2$-$\pi$-$\pi/2$) separated by a free fall time $T$. Here the momentum transfer is always in a single direction and the first $\pi/2$ pulse creates a superposition between the states $\left| a,p \right\rangle$ and $\left| b,p+ \hbar k_{e} \right\rangle$, where $a$ and $b$ refer to two hyperfine levels. We propose replacing this first pulse by a combination of a $\pi/2_+$ pulse and a $\pi_-$ pulse where the + and - represents the direction of the momentum transfer (Fig. \ref{fig:sequencepulses}). The first pulse creates a superposition similar to the one above and we tune the second pulse between the states $\left| b,p+ \hbar k_{e} \right\rangle$ and $\left| a,p+2 \hbar k_{e} \right\rangle$. In this way the superposition created has the form (up to small corrections that we consider later on)
\begin{equation}
\label{state}
\left| \Psi \right\rangle = \frac{1}{\sqrt{2}} \left( \left| p \right\rangle + \e^{i\sqrt{3} \pi /2} \left| p+2 \hbar k_{e} \right\rangle \right) \left| a \right\rangle,
\end{equation}
which is a superposition of two different momentum states in the same internal level. There is an extra phase that appears due to the nonresonant fields as we explain below.
We need to modify as well the other pulses needed to complete the gravimetric sequence (Fig. \ref{fig:sequencepulses}). The $\pi$ pulse is replaced by a  $\pi_+$, $\pi_-$ and $\pi_+$ pulses. The last $\pi/2$ pulse is replaced by a $\pi_-$ and $\pi/2_+$ pulses.

A similar strategy has been already demonstrated in Ref.~\cite{Berg2015} to measure the Earth's rotation rate using a symmetric momentum superposition with $\left| b,p-\hbar k_{e}\right\rangle$ and $\left| b,p+\hbar k_{e}\right\rangle$. This interferometer exhibits immunity to external magnetic field fluctuations. Their Raman beams come from independent phase-locked lasers with a fixed frequency. For atomic gravimetry, it becomes necessary to add a frequency ramp to compensate the Doppler shift of the free-falling atoms, and in addition, the proposed scheme requires alternating between ramps of opposite sign. The resulting modulation is quite complicated and it can be implemented in a simpler way using a fiber modulator. Using a modulator gives a simpler and rugged system, better suited for field applications. In this work, we study the effects induced on the interferometer by the additional laser lines of the laser modulation. We found that particular values of the Rabi frequency make the system less sensitive to imperfections.

\section{System Hamiltonian}
\label{Dynamics}

Consider an alkali-metal atom interacting with the two Raman pairs of section \ref{Intvel}. The atom is in free fall, and the beams drive transitions with opposite momentum transfer. The Hamiltonian of the system in the interaction picture is given by

\begin{align} \label{H}
    \begin{aligned}
    \hat{H}(t) &=\hat{H}_{0}+\hat{H}_{I}(t), \\
    \hat{H}_{0} &=\frac{\hat{p}^{2}}{2m}+mg\hat{z},\\
    \frac{\hat{H}_{I}(t)}{\hbar} &=\left( \Omega _{-}\e^{-ik_{e}\hat{z}} +\Omega _{+}\e^{ik_{e}\hat{z}} \right) \e^{i\alpha \left( t\right) } \hat{\sigma}_{+} \\
    &+\,\mathrm{C.\,C.}
    \end{aligned}
\end{align}

Here we do not consider the $x$ and $y$ components because they only contribute with a free expansion.
We used the rotating wave approximation in $\hat{H}_I$, with
$\hat{\sigma} _{\pm}$ the ladder operators between the hyperfine states $\ket{a}$ and $\ket{b}$. $\Omega _{\pm}$ are the Rabi frequencies of the Raman transitions, that for large detuning $\Delta$ with respect to the excited state $\ket{c}$, are given by~\cite{Arias17}

\begin{align} \label{Omeg}
    \begin{aligned}
    \Omega_{+}&=(\vec{E}_c\times\vec{E}_{+1}^*)\cdot\vec{M}_{+},\\
    \Omega_{-}&=(\vec{E}_{-1}\times\vec{E}_c^*)\cdot\vec{M}_{-},\\
    \vec{M}_{+}&=\frac{e^2}{4\hbar^2\Delta} \sum_c \bra{a}\vec{r}\ket{c}\times\bra{b}\vec{r}\ket{c}^*,\\
    \vec{M}_{-}&=\frac{e^2}{4\hbar^2\Delta} \sum_c \bra{b}\vec{r}\ket{c}\times\bra{a}\vec{r}\ket{c}^*.
    \end{aligned}
\end{align}

$\vec{E}_{-1}$, $\vec{E}_c$ and $\vec{E}_{+1}$ correspond to the electric field of the left sideband ($-1$), carrier ($c$) and the right sideband ($+1$) respectively. 
The effective wave vector $k_{e}$ has small deviations from $2|\vec{k}_c|$, that we neglected here but they have been studied in detail in Ref.~\cite{Carraz2012}. $\alpha \left( t\right) =\omega _{ba}t-\Xi \left( t\right) $, with $\omega _{ba}$ the hyperfine splitting and $\Xi(t)$ the integrated phase of the frequency difference $\Delta\omega = |\omega_{\pm1}(t)-\omega_c|$ of the two Raman beams in a pair.

We move to the reference frame fixed with the free-falling atom by means of a Galilean transformation (Appendix \ref{Galilean}), and the Hamiltonian \eqref{H} is transformed to

\begin{align} \label{H0TT2}
    \begin{aligned}
    \hat{H}^{'}_{0} &=\frac{\hat{p}^{2}}{2m}, \\
    \frac{\hat{H}^{'}_{I} (t)}{\hbar} &=\e^{-ik_{e}\hat{z}}\left( \Omega _{-}\e^{i\Theta
    _{-}\left( t\right) }\hat{\sigma}_{+}+\Omega _{+}^{\ast }\e^{-i\Theta_{+}\left( t\right) }\hat{\sigma}_{-}\right)\\&+\mathrm{C.C.,}
    \end{aligned}
\end{align}
where

\begin{equation}\label{OmegPM_def}
 \Theta_{\pm}\left(t\right)\equiv \alpha \left( t\right) \pm k_{e}z_{c}\left( t\right),
\end{equation}
and $z_c$ is the classical atom position given in Eq.~\eqref{varc}. With this transformation, we account for the gravitational term in the Hamiltonian. The price to pay is that the accelerated motion of the atom against the light fields introduces a phase (or frequency) shift encoded in $\Theta_{\pm}\left(t\right)$. Due to the difference in momentum transfer between $\Omega_+$ and $\Omega_-$, there is also a difference in the sign of this frequency shift, and this allows us to tune each transition separately. The transition selection is not perfect, and in what follows, we calculate the effect of the non-resonant contributions.

\section{Perturbative approach}
\label{DiffEqs}

\begin{figure}[H]
\centering
\includegraphics[width=80mm]{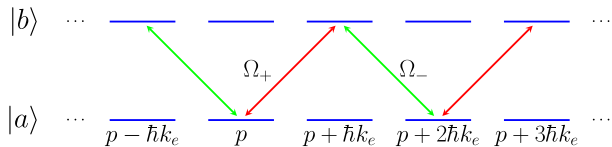}
\caption{States connected by the two Raman pairs.}
\label{fig:conectedstates}
\end{figure}

We expand the state of the atom in momentum space as

\begin{equation}\label{Gen_State_p}
 \left\vert \phi _{cm}(t)\right\rangle  =\int dp\,\e^{-i\frac{p^{2}t}{2m\hbar}}\left[ C^{a}_{p}\left(t\right)\left\vert
a\right\rangle +C^{b}_{p}\left(t\right) \left\vert b\right\rangle \right]
\left\vert p\right\rangle.
\end{equation}
Replacing the previous state into the Schr\"odinger equation for the free-falling atom (see Eq. \eqref{TScheq2}) and making use of $\exp(\mp ik_{e}\hat{z})=\int dp\left\vert p\right\rangle
\left\langle p\pm \hbar k_{e}\right\vert $ \cite{Moler1992} we get the following differential equation for the coefficients 

\begin{align} \label{dC}
    \begin{aligned}
    i\dot{C}^{b}_{p+n\hbar k_{e}}&=\Omega _{-}\e^{-i\left(\delta^{(n)}_pt+\omega_{R}t-\Theta_{-}\right)}C^{a}_{p+(n+1)\hbar k_{e}} \\
    &+\Omega_{+}\e^{i\left(\delta^{(n-1)}_p t+\omega_{R}t+\Theta_{+}\right)}C^{a}_{p+(n-1)\hbar k_e},\\
    i\dot{C}^{a}_{p+n\hbar k_e}&=\Omega_{-}^{\ast}\e^{i\left(\delta^{(n-1)}_p t+\omega_{R}t-\Theta_{-}\right)}C^{b}_{p+(n-1)\hbar k_{e}}\\
    &+\Omega_{+}^{\ast}\e^{-i\left(\delta^{(n)}_p t+\omega_{R}t+\Theta_{+}\right)}C^{b}_{p+(n+1)\hbar k_{e}},
    \end{aligned}
\end{align}
with $\delta^{(n)}_p=k_e (p+n\hbar k_{e})/m$ and $\omega_{R}=\hbar k_{e}^2/2m$ which is approximately four times the recoil frequency that has a value of 3.7~kHz in the case of $^{87}$Rb. The system couples an infinite chain of states separated by $\hbar k_e$ as shown in Fig.~\ref{fig:conectedstates}. The effective Hamiltonian corresponding to Eqs.~\eqref{dC} can be written as
\begin{widetext}
\begin{equation}\label{Heff_p_space}
\frac{\hat{H}(p,t)}{\hbar}=\Omega _{+}\e^{i(\Theta_{+}(t)+\omega_{R}t)}\sum_{n=-\infty}^{\infty}
\e^{i\delta^{(n)}_p t}\left\vert b,n+1 \right\rangle\!\!\left\langle a,n \right\vert+
\Omega _{-}\e^{i(\Theta_{-}(t)-\omega_{R}t)}\sum_{n=-\infty}^{\infty}
\e^{-i\delta^{(n-1)}_p t}\left\vert b,n-1 \right\rangle\!\!\left\langle a,n \right\vert+\mathrm{C.C.},
\end{equation}
\end{widetext}
where we have introduced the notation 

$\left\vert j,n\right\rangle=\left\vert j,p+n\hbar k_{e}\right\rangle$ with $j=a,b$. 

The Hamiltonian \eqref{Heff_p_space} is related to the one proposed in Ref.~\cite{Carraz2012}, but here we have neglected the small differences in momentum transfer of different Raman pairs, and instead we take into account the contributions from components with opposite momentum transfer.

Suppose the atom is initially prepared in the state $\left\vert a,0 \right\rangle$ and we want to make a transition with $\Omega_+$ to $\left\vert b,1 \right\rangle$. We achieve this by adding the appropriate frequency ramp ($\Xi(t)$) that compensates the detuning in Eq.~\eqref{delta}. In that case the argument of the exponential accompanying the $\Omega_+$ term in Eqs.~(\ref{dC},~\ref{Heff_p_space}) becomes zero, that is, $\omega_R t + \Theta_{+} = 0$ ($\delta_p^{(0)}$ is also zero for $p=0$), and ignoring the other non resonant terms we get resonant Rabi oscillations between the two levels. We separate the Hamiltonian \eqref{Heff_p_space} in two terms $\hat{H}=\hat{H_0}+\hat{V}$, with the first one containing the known evolution and the other the perturbations that oscillate at high frequency

\begin{align} \label{H0yV}
    \begin{aligned}
    \frac{\hat{H}_0}{\hbar}&=\Omega_{+}\sum_{n=-\infty}^{\infty}
    \e^{i\delta^{(n)}_p t}\left\vert b,n+1\right\rangle\!\!\left\langle a,n\right\vert+\mathrm{C.C.}, \\
    \frac{\hat{V}}{\hbar}&=\Omega_{-}\sum_{n=-\infty}^{\infty}
    \e^{-i\Delta^{(n-1)}_p t}\left\vert b,n-1\right\rangle\!\!\left\langle a,n\right\vert+\mathrm{C.C.},
    \end{aligned}
\end{align}
where $\Delta_p^{(n)}t=2k_{e}z_{c}(t)+2\omega_{R}t+\delta_p^{(n)}t$. We can write the evolution operator as a Dyson series

\begin{eqnarray}
    \label{DysonExpansion}
    \hat{U}(t,t_{0})&=&\hat{1}+\left(-\frac{i}{\hbar}\right)\int_{t_{0}}^{t}dt_{1}\hat{H}(t_{1})\\
    \notag
    &&+\left(-\frac{i}{\hbar}\right)^{2}\int_{t_{0}}^{t}dt_{1}\int_{t_{0}}^{t_{1}}dt_{2}\hat{H}(t_{1})\hat{H}(t_{2})+\cdots.
\end{eqnarray}

Writing the Hamiltonian as in Eqs.~\eqref{H0yV} we obtain the following series (Appendix \ref{Dysonpert})
\begin{widetext}
\begin{equation}
\label{GeneralPerturbationTerm}
\hat{U}(t)=\hat{U}_{0}(t)-\frac{i}{\hbar}\int_{0}^{t}dt_{1}\hat{V}(t_{1})\hat{U}_{0}(t_{1})+\sum_{n=2}^{\infty}\left(-\frac{i}{\hbar}\right)^{n}\int_{0}^{t}dt_{1}\cdots\int_{0}^{t_{n-1}}dt_{n}\prod_{k=1}^{n-1}\left[\hat{H}_{0}(t_{k})+\hat{V}(t_{k})\right]\hat{V}(t_{n})\hat{U}_{0}(t_{n}),
\end{equation}
\end{widetext}
where $\hat{U}_{0}(t)$ is the evolution operator of $\hat{H_0}$. Equation \eqref{GeneralPerturbationTerm} allows us to make a perturbative expansion of the evolution. The last term in the expansion is of second-order or higher in the perturbative parameters that we introduce below, and we neglect it from now on. The discussion is for the case when the $\Omega_{+}$ transition is resonant. An equivalent procedure can be followed straight forward when the $\Omega_{-}$ transition is the resonant one, as is the case for some of the other pulses in the sequence (See section \ref{pulse2}).

\section{First pulse}
\label{pulse1}

In this section, we analyze the first $\pi/2$ pulse in the sequence shown in Fig.~\ref{fig:sequencepulses}. The atoms are initially in the state $\left\vert a,0 \right\rangle$ and we want to create a superposition between this initial state and the state $\left\vert b,1 \right\rangle$ with the $\Omega_+$ field (red arrow in Fig.~\ref{fig:conectedstates}). The appropriate frequency ramp on the lasers is given by $\omega_R t + \Theta_{+} = 0$, as we explained before, and the evolution due to the $\Omega_+$ field is obtained from the first term of Eq.~\eqref{GeneralPerturbationTerm}. A residual constant detuning $\delta_p^{0}=k_e p/m$ appears in the transition coming from a momentum difference from the designed trajectory.
The velocity distribution in the $z$ axis must be narrow enough for this contribution to be small as quantified by the perturbative parameter

\begin{equation}
\label{betadef}
\beta=\frac{(k_ep/m)}{2\Omega},
\end{equation}
where in general $\Omega_{\pm}=\Omega\e^{i\theta_{\pm}}$ (up to small differences in the Rabi frequency due to the different detuning of the two Raman pairs) and we have considered the simplest case $\theta_{\pm}=0$.

The two states of interest $\left\vert a,0 \right\rangle$ and $\left\vert b,1 \right\rangle$ are also connected to the adjacent levels via the $\Omega_-$ field (green arrows in Fig.~\ref{fig:conectedstates}) perturbatively to first order in $\eta$ (See below). The contribution from these transitions is calculated from the second term of the evolution operator in Eq.~\eqref{GeneralPerturbationTerm}. We note here that these detuned transitions appear not only in our particular setup but in any setup with retro-reflected Raman beams and their effect on the signal must be taken into account. 

Since the $\Omega_-$ field has the opposite Doppler shift, the transition would be detuned by approximately $2 k_ev_z$ (Eq.~\eqref{delta}), with $v_z=\dot{z_c}$. We define a perturbative parameter

\begin{equation}
\label{etadef}
 \eta_{i}=\frac{\Omega}{2 k_e v_z(t_{i})},
\end{equation}
where $i$ refers to the pulse number in the sequence, and it is evaluated with the velocity at the end of that particular pulse.
The calculation for the first and second terms of Eq.~\eqref{GeneralPerturbationTerm} are shown in the Appendix~\ref{pulse1appendix} (considering $z_0=0$) and the wavefunction at the end of the pulse is given by

\begin{align} \label{solutionpulseone}
    \begin{aligned}
     \left\vert \psi_{1}\right\rangle& \simeq \frac{1}{\sqrt{2}}\left[1+i \left(1-\frac{\pi}{4}\right)\beta\right]\left\vert a,0\right\rangle\\
    &-\frac{i}{\sqrt{2}}\left(1+i\frac{\pi}{4}\beta\right)\left\vert b,1\right\rangle\\
    &-\frac{1}{\sqrt{2}}\left(\sqrt{2}-\Gamma_{\!\!\frac{\pi}{4}}^{\ast}\right)\eta_{1}\left\vert b,-1\right\rangle\\
    &+\frac{i}{\sqrt{2}}\e^{i(\pi\mu)}\Gamma_{\!\!\frac{\pi}{4}}\eta_{1}\left\vert a,2\right\rangle, 
    \end{aligned}
\end{align}
with $\Gamma_X=\exp{(iX/\eta)}$ and 
\begin{equation}\label{mudefinition}
    \mu=\frac{\hbar k_e^2}{2m\Omega} = \frac{\omega_R}{\Omega}.
\end{equation}
 The two perturbative states $\ket{b,-1}$ and $\ket{a,2}$ have a coefficient proportional to $\eta_{1}$, with $t_{1}=\tau_{1}=\pi/4\Omega$.

To validate Eq.~\eqref{solutionpulseone}, we implement a finite-difference numerical method to solve the coupled system of complex differential equations (Eq.~\eqref{dC}). We considered the coefficients with subindex $n=0,\pm2,\pm4,\pm6$, and set the remaining coefficients equal to zero for all times. 

\begin{figure}[H]
\centering
\includegraphics[width=80mm]{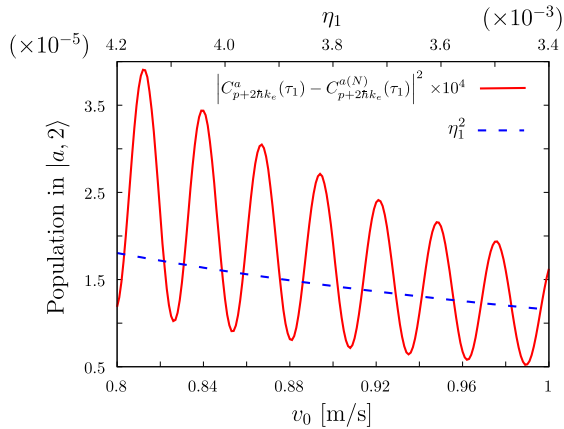}
    \caption{Population of state  $\ket{a,2}$ as a function of the initial velocity. The red-solid line represents the squared difference between the analytical and the numerical ($N$) solution at the end of the first pulse ($\tau_{1}$) multiplied by $10^4$. The blue-dashed line corresponds to $\eta_{1}^{2}$.}
    \label{fig:difsquare}
\end{figure}

Figure \ref{fig:difsquare} shows the modulus square of the difference between the analytical (Eq.~\eqref{solutionpulseone}) and numerical results. The agreement between the solutions improves as $v_{0}$ increases, and the deviation between the numerical and analytical solutions scales with a factor smaller than $\eta_{1}^{2}$, as expected for a perturbative expansion. 
\section{Second pulse}
\label{pulse2}

The second pulse in the sequence shown in Fig.~\ref{fig:sequencepulses} couples the states $\left\vert b,1 \right\rangle$ and $\left\vert a,2 \right\rangle$ with a $\pi$ pulse using the $\Omega_-$ field (green arrow in Fig.~\ref{fig:conectedstates}). The frequency chirp is such that $\Theta_-(t)-3\omega_R t=0$. The evolution is calculated in an equivalent way than for $\Omega_+$ but with $\Omega_+$ and $\Omega_-$ exchanging roles (Appendix \ref{pulse2appendix}).

\begin{eqnarray}
 \nonumber
 \ket{\psi_{2}}&\simeq&\frac{1}{\sqrt{2}}\e^{-i\mu\pi}\left[r_{a}^{+}+iq_{+}C_{d}\beta+i\eta_{2}\Gamma^{\ast}_{\frac{\pi}{4}}\right]\ket{a,0}\\
 \label{Equ:StateSplit}
 &-&\frac{1}{\sqrt{2}}\left(1+i\frac{5\pi}{4}\beta\right)
 \ket{a,2}\\
 \nonumber
 &-&\frac{1}{\sqrt{2}}\e^{i\mu\pi}\left[
 \e^{i\mu\pi}
 S_{d}\left(-i+p_{-}\beta\right)+wC_{d}\eta_{1}\right]\ket{b,-1}\qquad\\
 \nonumber
 &+&\frac{1}{\sqrt{2}}\e^{i\mu\pi}\Gamma_{\frac{\pi}{4}}\left[
 \eta_{1}+\left(1+\e^{i\mu\pi}\Gamma_{\frac{\pi}{2}}\right)\eta_{2}\right]\ket{b,1}\\
 \nonumber
 &+&\frac{1}{\sqrt{2}}\e^{i\mu(6\pi)}\Gamma_{\frac{3\pi}{4}}
 \eta_{2}\ket{b,3},
\end{eqnarray}

\noindent where $q_{+}=1+\frac{\pi}{4}$, $p_{-}=1-\frac{5\pi}{4}$,  $w=\sqrt{2}-\Gamma_{\frac{\pi}{4}}^{*}$, and
\begin{equation}
r_{a}^{+}=C_{d}+i
\e^{-i\mu\pi}
wS_{d}\eta_{1},
\end{equation}
being
\begin{align}
\begin{aligned}
C_{d}&=\cos\left(\frac{\pi\bar{\mu}}{2}\right),\\
S_{d}&=\frac{1}{\bar{\mu}}\sin\left(\frac{\pi\bar{\mu}}{2}\right),
\end{aligned}
\end{align}
with $\bar{\mu}^2=4\mu^2+1$ and $\eta_{2}$ is evaluated at $t=\tau_1+\tau_2$ and $\tau_{2}=\pi/2\Omega$. The field $\Omega_-$ couples the states $\left\vert b,1 \right\rangle$ and $\left\vert a,2 \right\rangle$ resonantly, but it also couples the states $\left\vert a,0 \right\rangle$ and $\left\vert b,-1 \right\rangle$ with a detuning close to 16 times the recoil frequency, introducing naturally the parameter $\mu$ (Eq. (\ref{mudefinition})).

Equation (\ref{Equ:StateSplit}) is a superposition of the states $\left\vert a,0 \right\rangle$ and $\left\vert a,2 \right\rangle$, among other spurious states. It gives the equivalent superposition of conventional atomic interferometers, except that now it is a superposition of momentum states with the same internal level. We notice that even in the case of $\beta\rightarrow0$ and $\eta_{i}\rightarrow0$ with $i=1,2$, the desired superposition is not perfect, and the state $\left\vert b,-1\right\rangle$ is still present. One way to minimize this problem is by reducing the Rabi frequency ($\Omega$) such that $\mu\gg1$ and $S_{d}\rightarrow0$, which is the approach followed in Ref.~\cite{Giese2013}. 
Another possibility is to make $S_{d}=0$ by setting  $\sin(\pi\bar{\mu}/2)=0$. This is the case for the following values of the Rabi frequency

\begin{equation}
\label{Eq:special_frequencies}
 \mu_m^2=\frac{1}{4}(4m^{2}-1),\qquad m=1,2,3,\cdots.
\end{equation}
With these special values of the Rabi frequency the population in $\left\vert a,0 \right\rangle$ makes multiple $2\pi$ transitions to the state $\left\vert b,-1 \right\rangle$ leaving no population in this last state. The state $\left\vert a,0 \right\rangle$ acquires a phase of $\e^{-i\pi\mu_m}$ in the process (Eq.~\eqref{state}). The first $(m=1)$ special Raby frequency for the case of $^{87}$Rb corresponds to a $\pi$ pulse of 30~$\mu$s, which is a typical pulse duration in gravimetry.

\begin{figure}[H]
\centering
\includegraphics[width=80mm]{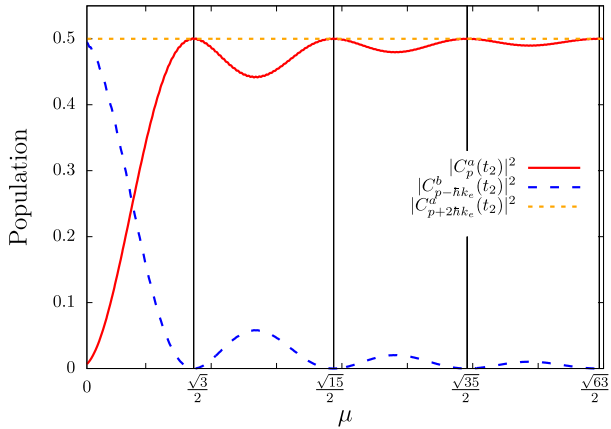}
    \caption{Populations of the different states after the second pulse (at time $t_{2}=\tau_{1}+\tau_{2}$) as a function of $\mu$ with an initial velocity of $v_{0}=5\ \text{m}/\text{s}$. The vertical lines indicate the special Rabi frequencies (Eq.~\eqref{Eq:special_frequencies}).}
    \label{fig:mu_variation}
\end{figure}

Figure (\ref{fig:mu_variation}) shows the population of the states $\ket{b,-1}$, $\ket{a,0}$ and $\ket{a,2}$ as we vary the parameter $\mu$. We observe that at special values of this parameter ($\mu_{m}$), the population of the state $\ket{a,0}$ is $1/2$ while the population of the state $\ket{b,-1}$ is close to zero. Using these special Rabi frequencies allows us to get the desired superposition of the states $\ket{a,0}$ and $\ket{a,2}$ by reducing the loss of population to spurious states like $\ket{b,-1}$.

\section{Interferometric fringes}
\label{fringesg}

\begin{figure}[H]
\centering
\includegraphics[width=80mm]{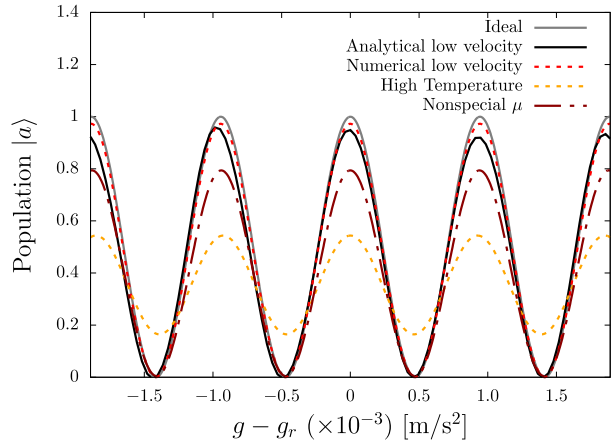}
    \caption{Interference fringes as we change the frequency chirp on the Raman beams. The solid light gray line corresponds to the ``Ideal" case of very low temperature ($T_{c}=0.14$ $\text{nK}$, corresponding to $\sigma_p=\hbar k_e/100$) and high initial velocity ($v_{0}=10\ \text{m}/\text{s}$). We show the effect of changing to a low initial velocity ($v_{0}=0.03\ \text{m}/\text{s}$) for the analytical calculation (solid blue line ``Analytical low velocity") and the numerical result (red dashed line ``Numerical low velocity"), or having a high temperature (yellow dashed line $T_{C}=1.45\ \mu\text{K}$) in the numerical calculation. The dotted dashed red line shows the reduction in visibility from the ``Ideal" case where special Rabi frequencies are not used (here we have $\mu=\sqrt{2}$).}
    \label{fig:Fringes}
\end{figure}

In this section, we calculate the result of the implementation of the seven pulses shown in Fig.~\ref{fig:sequencepulses}. We assume the special Rabi frequencies found in Eq.~\eqref{Eq:special_frequencies}, and we include corrections up to first order in $\beta$ and $\eta$. At the end of the interferometric sequence the population in the two hyperfine levels can be written as

\begin{eqnarray}
\nonumber
 \left|\braket{a|\psi_{F}}\right|^{2}&=&a_{0}+a_{0}^{\beta}\beta+a_{2}\eta_{2}+a_{3}\eta_{3}+a_{5}\eta_{5}+a_{6}\eta_{6}, \\
 \left|\braket{b|\psi_{F}}\right|^{2}&=&b_{0}+b_{0}^{\beta}\beta+b_{2}\eta_{2}+b_{3}\eta_{3}+b_{5}\eta_{5}+b_{6}\eta_{6},
 \qquad\quad\label{Eq::FinalPopulations}
\end{eqnarray}
where the zeroth order solutions 
\begin{align} \label{Probability_a,0}
    \begin{aligned}
     a_0&=\frac{1}{2}+\frac{1}{2}\cos\left(2k_{e}\gamma T^2\left[1+10\frac{\tau_1}{T}+\frac{95}{4}\frac{\tau_{1}^{2}}{T^2}\right]\right),\quad\\
     b_0&=\frac{1}{2}-\frac{1}{2}\cos\left(2k_{e}\gamma T^2\left[1+10\frac{\tau_1}{T}+\frac{95}{4}\frac{\tau_{1}^{2}}{T^2}\right]\right),
    \end{aligned}
\end{align}
give the expected fringes as shown in the ``Ideal" curve of Fig.~\ref{fig:Fringes}, where $\gamma=g-g_{r}$ and $g_{r}=(1/k_e) |\partial_t^2 \Xi(t)|$ is related to the frequency chirp applied to the sidebands $\omega_{\pm 1}(t)$. There is a factor of two in Eq.~\eqref{Probability_a,0}  because of the enhanced momentum separation in the superposition of momentum states. All of the other coefficients are given in Appendix \ref{coeficientsab}. The terms proportional to $\eta_1$, $\eta_4$ and $\eta_7$ give no contribution to the final signal. Some of the coefficients $a_i$ ($b_i$) have phases that might be difficult to control experimentally, but fortunately these coefficients have values of order $1$ and since they are multiplied by the perturbative parameter $\eta$, one can limit its contribution by making $\eta$ small.

A sample of laser-cooled atoms has a momentum distribution that can be well approximated by a Gaussian distribution with a standard deviation $\sigma_p$ centered at $p=0$ in the reference frame of the free-falling atom. There is no first-order correction in momentum $p$ ($\beta$) since the coefficients $a_0^\beta$ and $b_0^\beta$ (Eq. \eqref{Perturbative_ab_beta_0}) are equal to zero, and we would have corrections of second order in $\beta$ that depend on $\sigma_p$. We estimate the size of these corrections by extending the expansion to the second-order term in $\beta$. Equation~\eqref{Eq::FinalPopulations} is valid as long as this second-order correction is small, that is

\begin{equation}
    \label{Criterium_Beta}
    \frac{D k_{e}^{2}\sigma_p^{2}}{\left(2m\Omega\right)^{2}} < 1,
\end{equation}
where $D$ is the dimensionless second-order coefficient in the $\beta$ expansion of Eq.~\eqref{Eq::FinalPopulations}, that has an upper value of $D\approx6$.
Relating the momentum width ($\sigma_p$) to the temperature ($T_{C}$) through $\sigma_p=\sqrt{ mk_{B}T_{C}}$, with $k_{B}$ the Boltzmann constant, gives a temperature for the validity of our approach

\begin{equation}
    \label{TemperatureBound}
    T_{C} < \frac{4m\Omega^{2}}{ D k_{B} k_{e}^{2}},
\end{equation}
that for the case of $\text{Rb}^{87}$ gives $T_{C}< 0.32\ \mu \text{K}$. Even when the analytical solutions would not work well at higher temperatures, the numerical solution shows that there is still reasonable visibility at $T_C=1.45\ \mu\text{K}$ (corresponding to $\sigma_p=\hbar k_e$), as it is shown in the ``High Temperature" curve of Fig. \ref{fig:Fringes}. A temperature of $0.32\ \mu\text{K}$ is about 10 times smaller than what is obtained for a typical sample of laser-cooled rubidium atoms, and it is 800 bigger than what is obtained when a 1~ms velocity selection pulse is used, as these pulses are usually added for gravimetry. Hence, there is no need for a Bose condensed sample.

Figure~\ref{fig:Fringes} also shows the resulting fringes at a very low temperature with an atomic cloud with a small initial velocity. Here there is a nonnegligible contribution of the Raman pair that transfers momentum in the ``wrong" direction ($\eta$ corrections). The analytical expression (``Analytical low velocity" curve) gives results very similar to the numerical calculation (``Numerical low velocity curve") for a initial velocity as low as 0.03 m/s. The analytical results can be used to estimate the required minimal velocity of the atomic cloud in order to have smaller corrections than the desired precision. A free-fall time of 100~ms, for example, would give a typical velocity of 1~m/s that in the case of $^{87}$Rb and a $\pi$ pulse of 1~ms duration gives a parameter $\eta\sim 10^{-4}$. In the same figure we show the improvement obtained from using the special Rabi frequencies of Eq.~\eqref{Eq:special_frequencies}. In the ``nonspecial $\mu$" curve, we see a considerable reduction of the visibility when we do not take advantage of the special values of the Rabi frequency given in Eq.~\eqref{Eq:special_frequencies}.

\section{Conclusions.}

In this paper, we present an atomic interferometric scheme applied to gravimetry that prepares a superposition of two momentum states in the same internal level. This superposition becomes much less sensitive to external electromagnetic fluctuations and opens up new possibilities for interferometry, such as adding Bloch oscillations in the dark evolution time to increase the sensitivity~\cite{Xu19}. The scheme is implemented with a fiber modulator that transfers momentum in opposite directions and couples an infinite chain of momentum states.

We demonstrate a perturbative approach that allows us to identify a hierarchy in the transitions to other levels. Neighboring levels from the transition used for the gravimetry are coupled to first order, and subsequent levels are coupled to higher orders. We identify three perturbative parameters that determine the size of the corrections to the signal. The parameter $\eta$ is related to transitions having the opposite Doppler shift. The parameter $\mu$ is related to the ratio of the Rabi frequency with respect to four times the recoil frequency and quantifies the detuning for off-resonant transitions with the same sign of the Doppler shift. We find special values of the Rabi frequency that minimize the contribution from this parameter and improve the visibility of the interference fringes. The last parameter, $\beta$, is related to the temperature of the atomic cloud. This parameter contributes to the signal at second order for a symmetric velocity distribution.

We provide analytical expressions to calculate the effect of the undesired transitions coming from additional frequencies of the modulation process. It is easy to estimate the order of magnitude of the corrections based on the perturbative parameters. The scheme presented can be used to decide over experimental parameters required to reach a particular precision in gravitational measurements.

\section*{Acknowledgments}
We gratefully acknowledge funding from CONACyT Proyecto Fronteras 157, CB 254460, CB A1-S-18696, and FAI-UASLP. We thank M.A. Maldonado and W. M. Pimenta for valuable discussions.

\appendix

\section{Galilean transformation}
\label{Galilean}

We move to the reference frame fixed with the free-falling atom by means of a Galilean transformation~\cite{GalileanT} which consists of a momentum boost, a position translation and a phase shift
\begin{equation}
\label{G}
\hat{G}\equiv \exp\left(\int \frac{L_{c}}{%
\hbar }dt\right)\exp\left(-i\frac{\hat{p}z_{c}}{\hbar }\right)\exp\left(i\frac{p_{c}\hat{z}}{\hbar}\right),
\end{equation}
with $z_c$, $p_c$ the classical position and momentum and $L_c$ the Lagrangian along the classical trajectory
\begin{eqnarray}\label{varc}
\nonumber z_c&=&z_0 + v_0 t + \frac{1}{2}gt^2,\\
\label{pc}p_c&=&m (v_0 + g t),\\
\nonumber L_c&=&\frac{p_c^2}{2m}-mgz_c,
\end{eqnarray}
given in terms of the initial position $z_0$ and velocity $v_0$. Under this transformation the Hamiltonian \eqref{H} transforms as
\begin{eqnarray}\label{TH}
\nonumber \hat{G}^{\dag }\hat{H}_{0}\hat{G} &=&\frac{\left( \hat{p}+p_{c}\right) ^{2}}{2m}%
+mg\left( \hat{z}+z_{c}\right) , \\
\nonumber \frac{\hat{G}^{\dag }\hat{H}_{I}\hat{G}}{\hbar } &=&\left( \Omega _{-}\e^{-ik_{e}\left( 
\hat{z}+z_{c}\right) }+\Omega _{+}\e^{ik_{e}\left( \hat{z}+z_{c}\right) }\right)\e^{i\alpha \left( t\right) }
\hat{\sigma}_{+}\label{HIT} \\
&&+\mathrm{C.C.},
\end{eqnarray}%
where all the operators are now in the reference frame of the center of mass of the falling atom. The Schr\"odinger equation for $\ket{\psi}=\hat{G}\ket{\phi_{cm}}$ becomes
\begin{equation}
\label{TScheq}
i\hbar \partial _{t}\left\vert \phi _{cm}\right\rangle =\left( \hat{G}%
^{\dag }\hat{H}\hat{G}-i\hbar \hat{G}^{\dag }\partial _{t}\hat{G}\right)
\left\vert \phi _{cm}\right\rangle,
\end{equation}
where
\begin{equation}
\label{dG}
\hat{G}^{\dag }\partial _{t}\hat{G}=\frac{iL_{c}-i\dot{z}_{c}\left( 
\hat{p}+p_{c}\right) +i\dot{p}_{c}\hat{z}}{\hbar }, 
\end{equation}
here $\dot{z}_{c}=\partial _{t}z_c$ and $\dot{p}_{c}=\partial _{t}p_c$.  Equation \eqref{TScheq} becomes
\begin{equation}
\label{TScheq2}
i\hbar \partial _{t}\left\vert \phi _{cm}\right\rangle =\left( \hat{H}^{'}_{0}+\hat{H}^{'}_{I}\right) \left\vert \phi _{cm}\right\rangle, 
\end{equation}
with the Hamiltonian as in Eq.~\eqref{H0TT2} .

\section{Dyson series perturbative expansion}
\label{Dysonpert}

In general, an  evolution operator can be written as a Dyson series expansion of the form

\begin{eqnarray}
    \label{DysonExpansionAppen}
    \hat{U}(t,t_{0})&=&\hat{1}+\left(-\frac{i}{\hbar}\right)\int_{t_{0}}^{t}dt_{1}\hat{H}(t_{1})\\
    \notag
    &&+\left(-\frac{i}{\hbar}\right)^{2}\int_{t_{0}}^{t}dt_{1}\int_{t_{0}}^{t_{1}}dt_{2}\hat{H}(t_{1})\hat{H}(t_{2})+\cdots,
\end{eqnarray}
where $H(t_i)$ denotes the Hamiltonian of the system at different times.

The Hamiltonian of the system can be splitted into two parts $\hat{H}=\hat{H}_{0}+\hat{V}$ as defined by Eq.~\eqref{H0yV}, where $\hat{V}$ is the interaction that excites nonresonant transitions that would appear as small perturbations. The third term in the Dyson series (Eq. \eqref{DysonExpansionAppen}) would give the following four terms:

\begin{eqnarray}
    \label{Eq:first_reordenation}
    &&\left(-\frac{i}{\hbar}\right)^{2}\int_{t_{0}}^{t}dt_{1}\int_{t_{0}}^{t_{1}}dt_{2}\hat{H}_{0}(t_{1})\hat{H}_{0}(t_{2}), \\
    \label{Eq:second_reordenation}
    &&\left(-\frac{i}{\hbar}\right)^{2}\int_{t_{0}}^{t}dt_{1}\int_{t_{0}}^{t_{1}}dt_{2}\hat{V}(t_{1})\hat{H}_{0}(t_{2}), \\
    \label{Eq:third_reordenation}
    &&\left(-\frac{i}{\hbar}\right)^{2}\int_{t_{0}}^{t}dt_{1}\int_{t_{0}}^{t_{1}}dt_{2}\hat{H}_{0}(t_{1})\hat{V}(t_{2}), \\
    \label{Eq:fourth_reordenation}
    &&\left(-\frac{i}{\hbar}\right)^{2}\int_{t_{0}}^{t}dt_{1}\int_{t_{0}}^{t_{1}}dt_{2}\hat{V}(t_{1})\hat{V}(t_{2}).
\end{eqnarray}

Equation ~\eqref{Eq:first_reordenation} is part of the expansion of the operator $\hat{U}_{0}(t)$. We gather all these terms that do not contain $\hat{V}(t)$ to obtain $\hat{U}_{0}(t)$. In analogous form, we group Eq.~\eqref{Eq:second_reordenation} with all other terms of the Dyson expansion that contain $\hat{V}$ only once, such that it appears on the left, to get

\begin{align}
    \begin{aligned}
     &\int_{t_{0}}^{t}dt_{1}\hat{V}(t_{1})\left\{\hat{1}+\left(-\frac{i}{\hbar}\right)\int_{0}^{t_{1}}dt_{2}\hat{H}_{0}(t_{2})+\cdots\right\}\\
    &=\int_{t_{0}}^{t}dt_{1}\hat{V}(t_{1})\hat{U}_{0}(t_{1},t_{0}).
    \end{aligned}
\end{align}
Equation \eqref{Eq:fourth_reordenation} contains $\hat{V}$ twice and it will contribute at a higher order, so we combine it with all of the other higher order terms. We are left to analize Eq.~\eqref{Eq:third_reordenation}. It may seem
that this term is of first order in $\eta$ since $\hat{V}$ appears only once, but the position of the operator does matter. Here the operator $\hat{V}$ is on the right and it will be integrated twice, and on each integration the perturbative order increases, making this a second order term. Since $\hat{V}(t_2)$ is proportional to $e^{2ik_e z_c (t_2)}$, the integration on $t_2$ (Appendix \ref{pulse1appendix}) gives a term proportional to $\eta(t_1)e^{2ik_{e}z_{c}(t_{1})}$, and the second integration on $t_1$ would give a total factor of $\eta^{2}$.

Equation~\eqref{GeneralPerturbationTerm} shows the result of combining the terms according to their perturbative order, with the last term containing all the terms of order higher than one in all the parameters. It can be shown that this expansion is the solution of the time-dependent Schr{\"o}dinger equation.

\section{First pulse calculations}
\label{pulse1appendix}

The operator $\hat{U}_0(t)$ (Eq.~\eqref{GeneralPerturbationTerm}) can be written as follows

\begin{equation}
\label{FirstOrderOperator}
 \hat{U}_0(t)=\sum_{n=-\infty}^{\infty}\hat{U}^{(n)}_0(t),
\end{equation}
where each term in the sum corresponds to detuned Rabi oscillations between the levels $\left\vert a,n \right\rangle$ and $\left\vert b,n+1 \right\rangle$, that is

\begin{align} \label{OpertatorEvol_0}
    \begin{aligned}
    \hat{U}^{(n)}_0 &= f^{(n)}_p\left\vert a,n\right\rangle\!\!\left\langle a,n\right\vert + f^{(n)\ast}_p\left\vert b,n+1\right\rangle\!\!\left\langle b,n+1\right\vert
    \\
    &+ g^{(n)\ast}_p\left\vert a,n\right\rangle\!\!\left\langle b,n+1\right\vert - g^{(n)}_p\left\vert b,n+1\right\rangle\!\!\left\langle a,n\right\vert,
    \end{aligned}
\end{align}

where

\begin{align}
    \begin{aligned}
    f^{(n)}_p(t)&=\e^{-i\delta^{(n)}_p t/2}\cos\left(\tilde{\Omega}^{(n)}_{+}t\right)  
    \\
    &+i\e^{-i\delta^{(n)}_p t/2}\frac{\delta^{(n)}_p}{2\tilde{\Omega}^{(n)}_{+}}\sin\left(\tilde{\Omega}^{(n)}_{+}t\right),
    \qquad
    \\
    g^{(n)}_p(t)&=i\e^{i\delta^{(n)}_p t/2}\frac{\Omega^{\ast}_{+}}{\tilde{\Omega}^{(n)}_{+}}\sin\left(\tilde{\Omega}^{(n)}_{+}t\right) , \\
    \tilde{\Omega}^{(n)}_{+}&=\left[|\Omega_{+}|^2+\left(\delta^{(n)}_p/2\right)^2\right]^{\frac{1}{2}}. 
    \end{aligned}
\end{align}

In a similar way we separate the second term in Eq.~\eqref{GeneralPerturbationTerm} as
\begin{equation}\label{secondtermexpansion}
\left(-\frac{i}{\hbar}\right)\int_{t_{0}}^{t}dt_{1}\hat{V}(t_{1})\hat{U}_0(t_1)=\sum_{n=-\infty}^{\infty}\hat{U}_1^{(n)}(t),
\end{equation}
that involves transitions induced by $\Omega_-$ with

\begin{align} \label{OpertatorEvol_1}
    \begin{aligned}
    \hat{U}^{(n)}_1 & =  
    \chi^{(n-1)}_{(n)}\left\vert b,n-1 \right\rangle\!\!\left\langle a,n \right\vert
    \\
    &-\chi^{(n-1)\ast}_{(n-2)}\left\vert a,n \right\rangle\!\!\left\langle b,n-1 \right\vert \\
    &-\zeta^{(n-1)\ast}_{(n)}\left\vert b,n-1 \right\rangle\!\!\left\langle b,n+1 \right\vert\qquad
    \\
    &-\zeta^{(n-1)}_{(n-2)}\left\vert a,n \right\rangle\!\!\left\langle a,n-2 \right\vert,
    \end{aligned}
\end{align}
where
\begin{align} \label{Int2_OpratorEvol_Def1}
    \begin{aligned}
    \chi^{(n)}_{(m)}(t)&=-i\Omega_{-}\int_{t_{0}}^{t}dt_{1}\e^{-i\Delta^{(n)}_p t_1}f^{(m)}_p(t_1-t_{0}), \\
    \zeta^{(n)}_{(m)}(t)&=-i\Omega^{\ast}_{-}\int_{t_{0}}^{t}dt_{1}\e^{i\Delta^{(n)}_p t_1}g^{(m)}_p(t_1-t_{0}).
    \end{aligned}
\end{align}
It is convenient to do a time redefinition, so that the lower integration limit is always equal to zero. In this case we obtain

\begin{eqnarray}
\nonumber
 \chi^{(n)}_{(m)}(t)&=&-i\Omega_{-}e^{-i\Delta_{p}^{(n)}t_{0}}\int_{0}^{t-t_{0}}dt_{1}\e^{-i\tilde{\Delta}^{(n)}_p t_1}f^{(m)}_p(t_1),\label{Int1_OpratorEvol_Def2}\qquad  \\
 \zeta^{(n)}_{(m)}(t)&=&-i\Omega^{\ast}_{-}e^{i\Delta_{p}^{(n)}t_{0}}\int_{0}^{t-t_{0}}dt_{1}\e^{i\tilde{\Delta}^{(n)}_p t_1}g^{(m)}_p(t_1),\label{Int2_OpratorEvol_1}
\end{eqnarray}
where $\tilde{\Delta}_{p}^{(n)}t_{1}$ has the initial velocity recalibration $v_{0}\to v_{0}+gt_{0}$ included. The above integrals would be rapidly oscillating because they have a large detuning. The bigger the velocity the smaller the effect of this perturbation. We evaluate the integrals by the stationary phase method explained in \cite{Mandel}

\begin{align} 
    \begin{aligned}
     \chi^{(n)}_{(m)}(t)&\approx\left.\frac{\Omega_{-}\e^{-i\Delta_{p}^{(n)}t_{0}}\e^{-i\tilde{\Delta}^{(n)}_p t_1}f^{(m)}_p(t_1)}{2k_e v_z(t_{1})+2\omega_R+\delta_p^{(n)}+\delta_p^{(m)}/2}\right|_{0}^{t-t_{0}}, \\
    \zeta^{(n)}_{(m)}(t)&\approx\left.\frac{-\Omega^{\ast}_{-}\e^{i\Delta_{p}^{(n)}t_{0}}\e^{i\tilde{\Delta}^{(n)}_p t_1}g^{(m)}_p(t_1)}{2k_e v_z(t_{1})+2\omega_R+\delta_p^{(n)}+\delta_p^{(m)}/2}\right|_{0}^{t-t_{0}}.
    \end{aligned}
\end{align}
The denominator is dominated by $2 k_e v_z$ and the functions become proportional to $\eta$

\begin{align} \label{Int1_OpratorEvol_3}
    \begin{aligned}
    \chi^{(n)}_{(m)}(t)&\approx \left.\eta_{1} \e^{i -\left(\Delta^{(n)}_p t_0+\tilde{\Delta}_{p}^{(n)}t_{1}\right)}f^{(m)}_p(t_1)\right|_{0,\beta=0}^{t-t_{0}}, \\
    \zeta^{(n)}_{(m)}(t)&\approx\left.-\eta_{1} \e^{-i \left(\Delta^{(n)}_p t_0+\tilde{\Delta}^{(n)}_p t_{1}\right)}g^{(m)}_p(t_1)\right|_{0,\beta=0}^{t-t_{0}}.
    \end{aligned}
\end{align}

They are evaluated at $\beta =0$ since this is already a first order term in $\eta$. With these expressions we set limits on the magnitude of this term, altough its phase ($\Delta^{(n)}_p$) would show strong oscillations proportional to $1/\eta$. We apply the evolution operators of Eqs.~\eqref{FirstOrderOperator} and \eqref{secondtermexpansion} to the initial state $\left\vert a,0 \right\rangle$ with a $\pi/2$ pulse to get Eq.~\eqref{solutionpulseone} that has the two resonantly coupled states $\left\vert a,0 \right\rangle$ and $\left\vert b,1 \right\rangle$ and the spurious states $\left\vert b,-1 \right\rangle$ and $\left\vert a,2 \right\rangle$. All this calculation is for the case when the $\Omega_+$ transition is resonant. An analogous calculation can be done for the case when the $\Omega_-$ is the one that is resonant as is shown in the Appendix \ref{pulse2appendix}.

\section{Second pulse calculations}
\label{pulse2appendix}

In this appendix, we give the formulas for the case when the resonant transition is that driven by the field $\Omega_-$, along the same lines as Appendix C. We write

\begin{equation}
    \hat{U}_{0}(t)=\sum_{n=0}^{\infty}\hat{U}_{0}^{(n)}(t)
\end{equation}
with

\begin{align}
    \begin{aligned}
    \hat{U}_{0}^{(n)}&=h_{p}^{(n)}\ket{a,n+2}\bra{a,n+2}\\
    &+h_{p}^{(n)\ast}\ket{b,n+1}\bra{b,n+1} 
    \qquad\qquad
    \\
    &-j_{p}^{(n)\ast}\ket{b,n+1}\bra{a,n+2}
    \\
    &+j_{p}^{(n)}\ket{a,n+2}\bra{b,n+1}
    \end{aligned}
\end{align}
where
\begin{align}
    \begin{aligned}
     h_{p}^{(n)}(t)&=\e^{i\delta_{p}^{(n)}t/2}\cos\left(\tilde{\Omega}_{-}^{(n)}t\right) \\
    &-i\e^{i\delta_{p}^{(n)}t/2}\frac{\delta_{p}^{(n)}}{2\tilde{\Omega}_{-}^{(n)}}\sin\left(\tilde{\Omega}_{-}^{(n)}t\right)\\
    j_{p}^{(n)}(t)&=-i\e^{i\delta_{p}^{(n)}t/2}\frac{\Omega_{-}}{\tilde{\Omega}_{-}^{(n)}}\sin\left(\tilde{\Omega}_{-}^{(n)}t\right), 
    \qquad\qquad
    \\
    \tilde{\Omega}_{-}^{(n)}&=\left[\left|\Omega_{-}\right|^{2}+\left(\delta_{p}^{(n)}/2\right)^{2}\right]^{1/2}.
    \end{aligned}
\end{align}

The perturbative term is

\begin{equation}
    \left(-\frac{i}{\hbar}\right)\int_{t_{0}}^{t}dt_{1}\hat{V}(t_{1})\hat{U}_{0}(t_{1})=\sum_{n=-\infty}^{\infty}\hat{U}_{1}^{(n)}(t)
\end{equation}
with
\begin{align}
    \begin{aligned}
    \hat{U}_{1}^{(n)}&=\kappa_{(n-2)}^{(n)}\ket{b,n+1}\bra{a,n}\\
    &+\nu_{(n-2)}^{(n)}\ket{b,n+1}\bra{b,n-1}\qquad\qquad\\
    &+\kappa_{(n)}^{(n)\ast}\ket{a,n}\bra{b,n+1}\\
    &-\nu_{(n)}^{(n)\ast}\ket{a,n}\bra{a,n+2}
    \end{aligned}
\end{align}
where

\begin{align}
    \begin{aligned}
    \kappa_{(m)}^{(n)}(t)&=-i\Omega_{+}^{\ast}\int_{t_{0}}^{t}dt_{1}\e^{-i\Delta_{p}^{(n)}t_{1}}h_{p}^{(m)}(t_{1}-t_{0}), \\
    \nu_{(m)}^{(n)}(t)&=-i\Omega_{+}\int_{t_{0}}^{t}dt_{1}\e^{-i\Delta_{p}^{(n)}t_{1}}j_{p}^{(m)}(t_{1}-t_{0}).
    \end{aligned}
\end{align}

After a time redefinition similar to Eq.~\eqref{Int2_OpratorEvol_1} we get

\begin{eqnarray}
\nonumber
\kappa_{(m)}^{(n)}(t)&=&-i\Omega_{+}^{\ast}\e^{-i\Delta_{p}^{(n)}t_{0}}\int_{0}^{t-t_{0}}dt_{1}\e^{-i\tilde{\Delta}_{p}^{(n)}t_{1}}h_{p}^{(m)}(t_{1}), \qquad\quad\\
\nu_{(m)}^{(n)}(t)&=&-i\Omega_{+}\e^{-i\Delta_{p}^{(n)}t_{0}}\int_{0}^{t-t_{0}}dt_{1}\e^{-i\tilde{\Delta}_{p}^{(n)}t_{1}}j_{p}^{(m)}(t_{1}).
\end{eqnarray}
and using the stationary phase approximation we get

\begin{align}
    \begin{aligned}
    \kappa_{(m)}^{(n)}(t)&\approx\left.\frac{\Omega_{+}^{\ast}\e^{-i\Delta_{p}^{(n)}t_{0}}\e^{-i\tilde{\Delta}_{p}^{(n)}t_{1}}h_{p}^{(m)}(t_{1})}{2k_{e}v_{z}(t_{1})+2\omega_{R}+\delta_{p}^{(n)}+\delta_{p}^{(m)}/2}\right|_{0}^{t-t_{0}},\\
    \nu_{(m)}^{(n)}(t)&\approx\left.\frac{\Omega_{+}^{\ast}\e^{-i\Delta_{p}^{(n)}t_{0}}\e^{-i\tilde{\Delta}_{p}^{(n)}t_{1}}j_{p}^{(m)}(t_{1})}{2k_{e}v_{z}(t_{1})+2\omega_{R}+\delta_{p}^{(n)}+\delta_{p}^{(m)}/2}\right|_{0}^{t-t_{0}}.
    \end{aligned}
\end{align}
or

\begin{align}
    \begin{aligned}
    \kappa_{(m)}^{(n)}(t)&\approx\left.\eta_{2}\e^{-i\left(\Delta_{p}^{(n)}t_{0}+\tilde{\Delta}_{p}^{(n)}t_{1}\right)}h_{p}^{(m)}(t_{1})\right|_{0,\beta=0}^{t-t_{0}},\\
    \nu_{(m)}^{(n)}(t)&\approx\left.\eta_{2}\e^{-i\left(\Delta_{p}^{(n)}t_{0}+\tilde{\Delta}_{p}^{(n)}t_{1}\right)}j_{p}^{(m)}(t_{1})\right|_{0,\beta=0}^{t-t_{0}}.
    \end{aligned}
\end{align}

Using these equations, we obtained the state at the end of the second pulse as in Eq. (\ref{Equ:StateSplit}). The equations of Appendix C and D are recycled for the successive pulses to obtain the wave function of the whole experimental sequence.

\section{Perturbative coefficients $a_{i}$ and $b_{i}$.}
\label{coeficientsab}

In this appendix, we give the expressions for the coefficients of Eq.~\eqref{Eq::FinalPopulations}. The coefficients for $\ket{a}$ and $\ket{b}$ are:
\begin{eqnarray}
&&\left.\begin{split}\label{Perturbative_ab_beta_0}
 a_{0}^{\beta}&=0,\\
 b_{0}^{\beta}&=0,
\end{split}\right.\\
&&\left.\begin{split}\label{Perturbative_ab_2}
 a_{2}&=-\cos\left(\Phi_{g}\right)\sin\left(\phi_2\right),\\
 b_{2}&=-\sin\left(\Phi_{g}\right)\cos\left(\phi_2\right),
\end{split}\right.\\
&&\left.\begin{split}\label{Perturbative_ab_3}
 a_{3}&=-\cos\left(\Phi_{g}\right)\left[\sin\left(\phi_x\right)+\sin\left(\phi_y\right)-\sin\left(\phi_z\right)\right],\qquad\\
 b_{3}&=\sin\left(\Phi_{g}\right)\left[\cos\left(\phi_x\right)+\cos\left(\phi_y\right)-\cos\left(\phi_z\right)\right],
\end{split}\right.\\
&&\left.\begin{split}\label{Perturbative_ab_5}
    a_{5}&=\cos\left(\Phi_{g}\right)\sin\left(\phi_5\right),\\
    b_{5}&=-\sin\left(\Phi_{g}\right)\cos\left(\phi_5\right),
\end{split}\right.\\
&&\left.\begin{split}\label{Perturbative_ab_6}
    a_{6}&=\cos\left(\Phi_{g}\right)\sin\left(\phi_6\right),\\
    b_{6}&=-\sin\left(\Phi_{g}\right)\cos\left(\phi_6\right),
\end{split}\right.
\end{eqnarray}

\noindent with 
\begin{eqnarray}
 \nonumber\Phi_g &=& k_e\gamma T^2,\\
 \nonumber\phi_{x}&=&\Phi_{g}\xi_{x}+2k_{e}v_{0}T\Lambda_{x}+k_{e}gT^2\Lambda_{x}^2+\frac{\mu}{2}(5\pi+4\Omega T),\\
 \nonumber\phi_{y}&=&\Phi_{g}\xi_{y}+2k_{e}v_{0}T\Lambda_{y}+k_{e}gT^2\Lambda_{y}^2+\mu(\pi+2\Omega T),\\
 \nonumber\phi_{z}&=&\Phi_{g}\xi_{z}+2k_{e}v_{0}T\Lambda_{z}+k_{e}gT^2\Lambda_{z}^2+\frac{\mu}{2}(5\pi+4\Omega T),\\
 \phi_2 &=& \Phi_g+2k_{e}v_{0}\tau_1+k_{e}g\tau_1^2,\\
 \nonumber\phi_5&=&\Phi_{g}\xi_5-2k_{e}v_{0}T\Lambda_5-k_{e}gT^2\Lambda_5^2-2\mu(2\pi+\Omega T),\\
 \nonumber\phi_6&=&\Phi_{g}\xi_6-4k_{e}v_{0}T\Lambda_6-4k_{e}gT^2\Lambda_6^2-\mu(5\pi+4\Omega T),\qquad
\end{eqnarray}

\noindent where the dimensionless numbers are given by

\begin{align}
\begin{aligned}
 \xi_x &= \frac{3}{2}\left(1+\frac{26}{3}\frac{\tau_1}{T}+\frac{111}{6}\frac{\tau_1^2}{T^2}\right),\\
 \xi_y &= \xi_x,\\
 \xi_z &= \frac{3}{2}\left(1+\frac{20}{3}\frac{\tau_1}{T}+\frac{31}{2}\frac{\tau_1^2}{T^2}\right),\\
 \xi_5 &= \frac{5}{2}\left(1+\frac{46}{5}\frac{\tau_1}{T}+\frac{211}{10}\frac{\tau_1^2}{T^2}\right),\\
 \xi_6 &= \left(1+12\frac{\tau_1}{T}+\frac{147}{4}\frac{\tau_1^2}{T^2}\right),\\
 \Lambda_x &= 1+5\frac{\tau_1}{T},\\
 \Lambda_y &= 1+3\frac{\tau_1}{T},\\
 \Lambda_z &= \Lambda_x,\\
 \Lambda_5 &= 1+7\frac{\tau_1}{T},\\
 \Lambda_6 &= 1+\frac{9}{2}\frac{\tau_1}{T}.
\end{aligned}
\end{align}

\noindent $T$ is much larger than any light-pulse duration as it is commonly assumed. 

\end{document}